\documentclass[%
 reprint,
superscriptaddress,
 amsmath,amssymb,
prl
]{revtex4-1}

\usepackage{graphicx}
\usepackage{subfigure}
\usepackage{dcolumn}
\usepackage{bm}
\usepackage{color}

\def\nin{\noindent} 
\def\beq{\begin{equation}}
\def\eeq{\end{equation}}

\begin{document}

\preprint{APS/123-QED}

\title{On universal structural characteristics of granular packs}

\author{Takashi Matsushima}
 \email{tmatsu@kz.tsukuba.ac.jp}
\affiliation{%
 Department of Engineering Mechanics and Energy, University of Tsukuba, Tsukuba, Japan\\
}%


\author{Raphael Blumenfeld}%
 \email{rbb11@cam.ac.uk}
\affiliation{
Earth Science and Engineering, Imperial College London, London SW7 2AZ, UK\\
}%
\affiliation{
Cavendish Laboratory, Cambridge University, JJ Thomson Avenue, Cambridge CB3 0HE, UK
}%


\date{\today}

\begin{abstract}

\nin Dependence of structural self-organization of granular materials on preparation and grain parameters is key to predictive modelling. We study 60 different mechanically equilibrated polydisperse disc packs, generated numerically by two protocols. We show that, for same-variance disc size distributions (DSDs):  
1. the mean coordination number of rattler-free packs vs. the packing fraction is a function independent of initial conditions, friction and the DSD: 
2. all quadron volume and cell order distributions collapse to universal forms, also independent of the above. 
We conclude that, contrary to common wisdom, equilibrated granular structures are determined mainly by the packing protocol and higher moments of the DSD.

\end{abstract}

\pacs{Valid PACS appear here}
\maketitle

\nin Self-organization of granular systems into solids is of primary importance in science and engineering \cite{EdOa89a, Vogel_Roth_2003, Cheng_etal_1999, BaBl02, Aste_etal_2007, Song-etal2008}.
A central quest in the field is for relations between the physical properties of mechanically stable many-grain packs and their \text{red}{emergent} structural characteristics. Yet, there is so far no systematic way quantify such relations. Moreover, different structural properties often correlate well with different physical properties. For example, void size distribution and connectivity correlate well with permeability to flow, which is relevant to underground water, pollutant dispersion and oil extraction \cite{Vogel_Roth_2003}; catalysis and heat exchange between the grains and a fluid in the void space are more sensitive to the solid-void surface distribution \cite{Cheng_etal_1999}; and the mechanics of granular solids is governed by an interplay between the structure and the force transmission through the intergranular contacts \cite{BaBl02}.
The problem is complicated by the many observations that the structural characteristics depend on the distributions of grain shapes, sizes, intergranular friction coefficients and preparation history. 
A key to progress is the identification of emergent structural properties that are independent of some of these parameters. 

\nin The aim of this paper is to identify such universality and show that, contrary to current belief, the distributions of a number of structural features of random packs of planar (2D) granular solids collapse to common forms for all intergranular friction coefficients, a wide range of initial preparation states, at least two preparation protocols, and several same-variance grain size distributions. 

\begin{figure}["here"]
\includegraphics[width=0.25\textwidth]{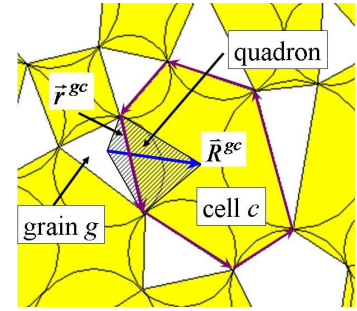}
\caption{The quadron (shaded) is the quadrilateral whose diagonals are vectors $\vec{R}^{gc}$ and $\vec{r}^{gc}$, defined in the text.}
\label{Fig01_quadron}
\end{figure}

\nin We analyse the structure of numerically generated and mechanically equilibrated 2D polydisperse disc packs, using the quadron-based structural description\cite{BaBl02,BlEd03,BlEd06}. In this method, illustrated in Fig. \ref{Fig01_quadron}, one first defines the centroids of grains (discs) $g$ and cells $c$ as the mean position vectors of the contact points around them, respectively. 
Next, the contact points around every disc are connected to make polygons, whose edges are vectors, $\vec{r}^{gc}$, that circulate the grain in the clockwise direction. 
Then, one extends vectors $\vec{R}^{gc}$ from the disc centroids $g$ to the cell centroids $c$ that surround them. A quadron is the quadrilateral whose diagonals are the vectors  $\vec{r}^{gc}$ and  $\vec{R}^{gc}$. 
In the absence of body forces, equilibrated cells are convex and, for convex grains, the pairs $\vec{r}^{gc}$ and $\vec{R}^{gc}$ generically intersect. For brevity, we index the vectors defining a quadron by $q$ rather than $gc$. 
The quadron's shape is  quantified by the structure tensor $C^q = \vec{r}^q\otimes\vec{R}^q$
and its volume is $V_q= \frac{1}{2}\mid\vec{r}^q \times \vec{R}^q\mid$ providing a quantitative measure of the local structure. 

\nin This description is convenient for measuring local structural characteristics and, as such, has a significant advantage over methods that average over arbitrarily-defined structural properties. The tessellation by quadrons is also preferable to Voronoi-based tessellations\cite{Vor,Vogel_Roth_2003,Cheng_etal_1999,Song-etal2008} both because it makes possible an unambiguous local tensorial description and because it preserves the connectivity information.

\nin For our numerical experiments we used the Discrete Element Method \cite{Cundall1979,Mat-Chang-2011}, in which disc motions follow Newton's second law with an incremental time marching scheme. For the interaction between discs, we use a repelling harmonic interaction potential, characterised by normal and tangential spring constants, $k_n$ and $k_t$, respectively, activated on contact and overlap between discs. Without loss of generality, we set $k_t/k_n=1/4$ in this study.

\nin Since crystalline structures are well understood, we focus on fully disordered systems and study
two disc size distributions (DSDs). One is log-normal, (LN), due to its wide use in civil engineering and soil sciences\cite{Mitchell-Soga-2005}, $P (D) = \exp [ ( \ln D - \ln D_0 )^2 / 2 \sigma^2 ] / \sqrt{2 \pi} \sigma D$, with $D_0=1.0$ and $\sigma=0.2$, whose mean disc size and mode are, respectively, $\bar{D}=1.02$ and $D_{mode}=0.961$. The other is uniform, (U), $0.663 \le D \le 1.38$, having the same mean and variance. 

\nin Our granular systems are generated as follows. 
First, we construct several random packs in a double periodic domain, engineered to be initially on the verge of jamming. The packs consist of $21400 \pm 1000$ discs and are made at packing fractions $\phi=0.72, 0.76, 0.82$ and $0.84$ for both the LN and U systems.
 The small variation in the numbers of particles is due to the different densities of the initial configurations.
These very loose-, loose-, intermediate- and dense configurations (respectively, VLIS, LIS, IIS and DIS) are used as initial states for a subsequent packing procedure as follows. 
All the discs are assigned a friction coefficient $\mu$ and the systems are then packed by two protocols: a slow isotropic compression (ISO) for LN and U systems and an anisotropic compression (ANISO) for LN systems. This is done by changing the periodic length either in both directions or only in one, such that the ratio of mean disc overlap to $\bar{D}$ is $10^{-5}$.
No gravitational force is used and the compression continues until the fluctuations of disc positions (per mean disc diameter) and intergranular forces (per mean average contact force) are below very small thresholds - $10^{-9}$ and $10^{-6}$, respectively. This procedure is carried out from each initial state for five different values of $\mu$: $0.01, 0.1, 0.2, 0.5$ and $10$, giving altogether 60 different systems: 40 of LN DSDs and 20 of U DSDs.

\begin{figure}["here"]
\begin{minipage}[t]{0.45\textwidth}
\includegraphics[width=1.0\textwidth]{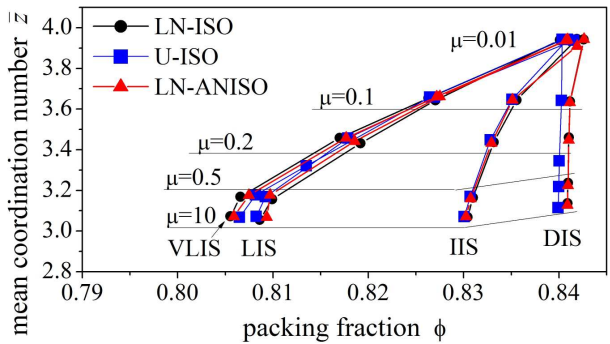}
\caption{Mean coordination numbers vs. packing fractions for all our systems. Different initial states VLIS, LIS, IIS DIS) and disc friction coefficients ($\mu$=0.01, 0.1, 0.2, 0.5, 10) give distinctly different final jammed packs.
}
\label{Fig02_z-phi}
\end{minipage}
\end{figure} 

\nin For every system, we computed the packing fraction, the mean coordination number, and a range of structural properties to be described below. 
For the determination of $\bar{z}$, we disregarded `rattlers', i.e. discs with one or no force-carrying contact. The dependences of $\bar{z}$ on $\phi$ are shown in Fig. \ref{Fig02_z-phi}. 
The upper and lower bounds, $\bar{z}_{max}=4$  and $\bar{z}_{min}=3$, correspond generically to isostatic states of smooth ($\mu=0$) and frictional ($\mu\to\infty$) discs, respectively.
The five systems with $\mu=0.01$ settle into states that are very close to the ideal frictionless jammed state. As $\mu$ increases the packs converge to increasingly different final states in the $\bar{z}$-$\phi$ plane, as shown in Fig.  \ref{Fig02_z-phi}.

\begin{figure}["here"]
\begin{minipage}[t]{0.45\textwidth}
\includegraphics[width=1.0\textwidth]{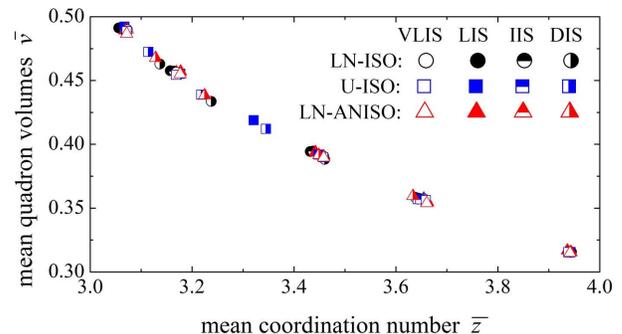}
\caption{The mean quadron volume $\bar{v}$ vs. the mean coordination number $\bar{z}$. The graph collapses for all inter-granular frictions, all initial states and both DSDs.}
\label{Fig03_v-z}
\end{minipage}
\end{figure}

\nin To study the mean quadron volumes, it is convenient to normalize them by the mean disc volume after removing rattlers  $\bar{V}_g'$, $\bar{v}\equiv V_q/ \bar{V}_g'$. 
In Fig. \ref{Fig03_v-z}, we plot $\bar{v}$ against $\bar{z}$ for all the systems and discover that all the points fall nicely on one curve. 

\begin{figure}["here"]
\begin{minipage}[t]{0.45\textwidth}
\includegraphics[width=1.0\textwidth]{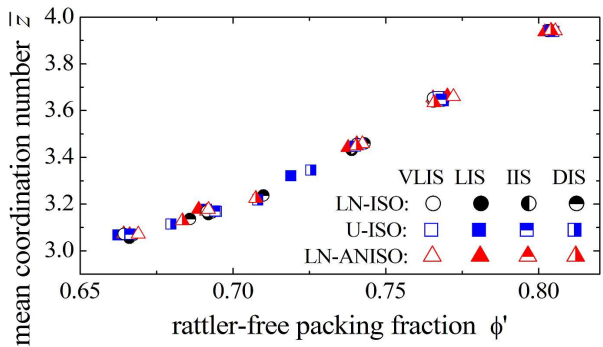}
\caption{The collapse of the relation between mean coordination number and rattlers-free packing fraction. }
\label{Fig04_z-phir}
\end{minipage}
\end{figure}

\nin As we shall see below, discarding the rattlers is essential to understanding this collapse. 
The total volume and the rattler-free solid volume are, respectively, $V=\sum_{q=1}^{N_q} V_q\equiv N_q\bar{V}_q$ and $V'_s=\sum_{g=1}^{N'_g} V'_g$, where $N_g$ and $N'_g$ are the total number of discs with and without rattlers, respectively, and $V'_g$ is the volume of (non-rattler) disc $g$. The rattler-free packing fraction is then $\phi'=N'_g \bar{V}'_g/ N_q\bar{V}_q$. Recalling that the total number of quadrons is $N_q=N'_g\bar{z}$ and $\bar{v}\equiv \bar{V}_q/\bar{V}'_g$, we obtain

\beq
\phi'=\frac{1}{\bar{v}\bar{z}}
\label{eq:model03}
\eeq 
Based on (\ref{eq:model03}) and the collapse in Fig. \ref{Fig03_v-z}, we expect $\phi'$ to be a function of $\bar{z}$ alone, which is indeed confirmed in Fig. \ref{Fig04_z-phir}. Note that this collapse is in contrast to current wisdom \cite{Smetal29, Song-etal2008}.
As seen from Figs. \ref{Fig04_z-phir} and \ref{Fig02_z-phi}, the rattler-free packing fractions are considerably smaller, but it is the latter packing fractions that are commonly reported in the literature. Significantly, the collapse of the plots in Fig. \ref{Fig02_z-phi} shows that the differences for different initial state is only due to the different rattler fractions.
It is also interesting that, while the curves of Fig. \ref{Fig02_z-phi} are sub-linear, their collapsed rattler-free form is super-linear. 

\begin{figure}["here"]
\begin{minipage}[t]{0.45\textwidth}
\includegraphics[width=1.0\textwidth]{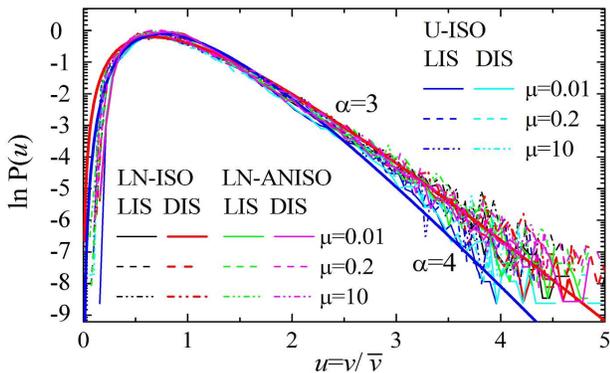}
\\*[-8pt] 
\caption{The collapse of the PDFs of the normalized quadron volumes, $u\equiv v/\bar{v}$, for all the systems, compared with two $\Gamma$ distributions (Eq.(\ref{GammaDist})), with $\alpha=3$ and $4$.
}
\label{Fig05_lnP-u}
\end{minipage}
\end{figure} 

\nin This collapse prompted us to consider more closely the probability density functions (PDFs) of the quadron volumes. 
Since the mean quadron volume increases with $\mu$, we scale the quadron volumes by their means, $u\equiv  V_q / \bar{V}_q=v/\bar{v}$. We find that this is sufficient to collapse the PDFs of all the systems almost perfectly onto one curve {\it independent of friction, initial state, the protocol used and the DSD used} (Fig. \ref{Fig05_lnP-u}).  

\nin Our best fit to the collapsed curve is the $\Gamma$ distribution 

\beq
P(u) = \frac{\alpha^\alpha}{\Gamma(\alpha)} u^{\alpha-1} e^{-\alpha u} 
\label{GammaDist}
\eeq
with $\alpha$ between 3 and 4.
These observations have several implications. (i) The quadron description makes possible to collapse the statistics of all the systems, thus providing better insight into the fundamental characteristics of granular packs. (ii) The structural characteristics are determined by physical mechanisms that transcend the effects of friction, initial state and some aspects of the DSD, which can all be scaled away. (iii) The collapse is only possible when ratters are disregarded, suggesting an inherent relation between the self-organization of the structure and the force-carrying backbone during the packing process. 

\nin To explore the origin of the above collapses, we use a recently-proposed decomposition of the quadron volumes into conditional PDFs\cite{Fretal08}

\beq
P\left(v\right)=\sum_e eQ(e) P\left(v\mid e\right)
\label{eq_decomposition}
\eeq
where $Q(e)$ is the occurrence probability of cells of order $e$ and $P\left(v\mid e\right)$ is the conditional PDF of the normalized quadron volume, given that it belongs to a cell of order $e$.
Frenkel et al.\cite{Fretal08} argued, on the basis of geometrical considerations, that $P\left(v\mid e\right)$ should be independent of intergranular friction. Their argument was that, given a collection of $N$ arbitrary grains, the number of ways to arrange $e$ grains into a cell of order $e$  depends only on the grains shapes. 
By and large, our results seem to support to this argument. However, we observe  a small, but systematic, $\mu$-dependence of $P\left(v\mid e\right)$ for $e> 6$. We believe that this is due to effects of mechanical stability on cell shapes. This issue is tangential to the thrust of this work and will be discussed elsewhere.

\begin{figure}["here"]
\begin{minipage}[t]{0.45\textwidth}
\includegraphics[width=1.0\textwidth]{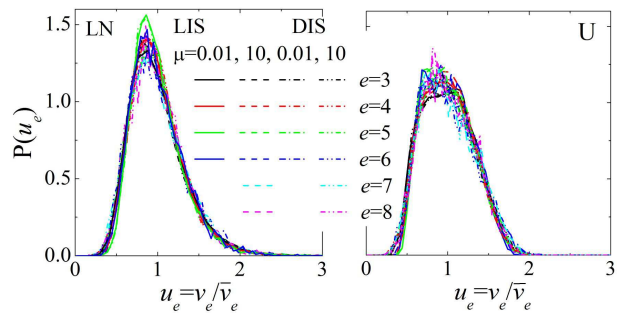}
\\*[-8pt] 
\caption{The collapse of the conditional PDFs of the normalized quadron volumes for all the systems for log-normal (left) and uniform (right) DSD.}
\label{Fig6-Pue}
\end{minipage}
\end{figure} 

\nin Significantly, we find that friction and initial states can also be scaled away for $P\left(v\mid e\right)$ -- all the conditional PDFs collapse under scaling the quandron volumes by the mean $v_e \equiv \int v P\left(v\mid e\right) dv$! 
The collapsed forms exhibit distinct differences for the different DSDs (Fig. \ref{Fig6-Pue}): the flatness and the finite support of the uniform DSD give rise to a flatter maximum and a tail cutoff. 
The tail dependence on the DSD suggests that $P\left(v\mid e\right)$ cannot have a universally exponential tail. Combined with (\ref{eq_decomposition}), this means that the observed exponential tail of $P(v)$ {\it must} originate in $Q(e)$.

\begin{figure}["here"]
\begin{minipage}[t]{0.45\textwidth}
\includegraphics[width=1.0\textwidth]{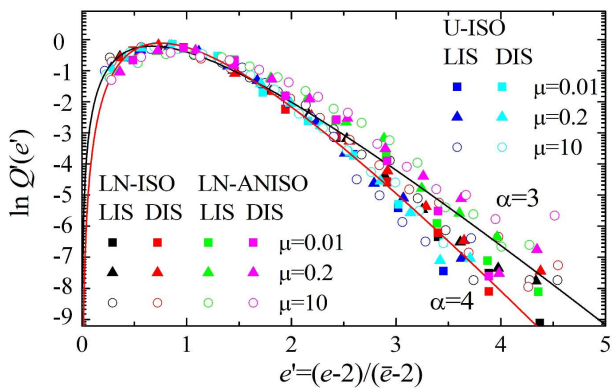}
\caption{The apparent collapse of all the PDFs of normalized cell orders, $Q'(e')$, $e'=(e-2)/( \bar{e}-2 )$, for log-normal and uniform DSD, compared with two $\Gamma$ PDFs with $\alpha=3$ and $4$.}
\label{Fig7Qen}
\end{minipage}
\end{figure}

\nin The PDF $Q(e)$ is central to understanding the structure\cite{BlToMaSoon}. 
Its mean, $\bar{e}$, is related to $\bar{z}$ as follows\cite{Fretal08}. 
Consider the system as a graph whose vertices are the $N$ disc centres, its edges are the lines connecting discs in direct contact and each face made by these edges corresponds to one of the $N_C$ cells. Each edge corresponds to a contact point, whose number is $N_E=N\bar{z}/2$. This graph satisfies Euler's relation in the plane $N-N_E+N_C = 1$ and, on substituting $N_E$, we get $N_C = N(\bar{z}/2-1) + 1$. The mean cell order, $\bar{e}$, is the mean number of edges per cell and, since each edge is shared between two cells, we have $\bar{e} = 2N_E/N_C$. It follows that

\begin{eqnarray}
\bar{e} = \frac{2 \bar{z}}{ \bar{z} -2} + O\left(\frac{1}{\sqrt{N}}\right) ; \bar{z} = \frac{2 \bar{e}}{ \bar{e} -2} + O\left(\frac{1}{\sqrt{N}}\right) 
\label{eq:mean_e}
\end{eqnarray}
where the rightmost terms are boundary corrections, whose negligibility we verified in our numerical systems. 
$Q(e)$ depends on the friction through its sensitivity to the value of $\bar{z}$, while $P\left(v\mid e\right)$ has been shown to be hardly dependent on $\mu$\cite{Fretal08}. Using insight from relation (\ref{eq:mean_e}), we find that all the $Q(e)$ curves collapse when plotted as $Q\left(e'=\frac{e-2}{\bar{e}-2}\right)$ for both the log-normal and uniform DSDs (Fig. \ref{Fig7Qen}).
Both the collapsed PDFs are fit well by a $\Gamma$ distribution of the form (\ref{GammaDist}), with $\alpha$ between 3 and 4, which is similar to the value observed for $P(u)$. This may suggest that the exponential tail of $P(u)$ is dominated by $Q(e)$ and is much less affected by $P\left(v\mid e\right)$. 

\nin To conclude, we have studied the effects of intergranular friction, initial state, disc size distribution and change in packing protocol, on the structural characteristics that disordered 2D granular systems pack into.
First, we demonstrated that, for our systems, there is a universal relation between the the mean coordination number and the rattler-free packing fraction. This relation is independent of the intergranular friction coefficient, the initial state of the packing process, the form of two same-variance DSDs and of which of two packing protocols we used, all of which have been believed previously to affect strongly this relation. Since the rattler-free structure carries the internal stress, this finding suggests that the universality is a result of the self-organization directly related to the evolution of the force distribution in the packing process. 
Next, we used the quadron description to quantify the grain-scale structure by focusing on the statistics of two variables: the quadron volumes $v$ and the cell order $e$. 
We found that the total and conditional distributions of these variables can be scaled to collapse to universal forms. These forms appear to be also independent of the above parameters. 
While we are convinced that such a collapse can always be found and that it should be independent of 
friction and initial states, we cannot rule out the possibility that a systematic exploration of the self-organization under a wide range of possible packing processes and DSDs might lead to different collapsed forms. It is possible that the collapse is universal only to classes of processes and DSDs and further work is required in this direction.
We also found that the tail of the quadron volume PDF is exponential, potentially supporting the conjecture of an exponential Boltzmann-factor-like weight of microstates in the statistical mechanical description of granular systems\cite{EdOa89a,BlEd03,HiBl12}. We demonstrated that the exponential tail most likely originates in $Q(e)$.

\nin This study focused on understanding the limit structural characteristics of purely disordered systems because the structures of crystalline systems are straightforwardly quantifiable. We believe that our results form an important step toward a comprehensive understanding of the self-organization of structures of mixtures of ordered-disordered granular packs.

\nocite{*}

 \bibliography{cell_geometry}

\end{document}